\documentclass[10pt,journal,compsoc]{IEEEtran}
\usepackage{cite}
\usepackage{amsmath,amssymb,amsfonts}
\usepackage{graphicx}
\usepackage{textcomp}
\usepackage{xcolor}
\usepackage{caption}
\usepackage{subcaption}
\usepackage{dirtytalk}
\usepackage{hyperref}
\usepackage{multirow,verbatim}
\usepackage{array}
\usepackage{color, colortbl}
\usepackage{soul}
\usepackage{tikz}
\usepackage{algorithm}
\usepackage{algpseudocode}
\usepackage{enumitem}
\usepackage{arydshln}
\usepackage{epsfig,endnotes}

\newcolumntype{L}[1]{>{\raggedright\let\newline\\\arraybackslash\hspace{0pt}}m{#1}}
\newcolumntype{C}[1]{>{\centering\let\newline\\\arraybackslash\hspace{0pt}}m{#1}}
\newcolumntype{R}[1]{>{\raggedleft\let\newline\\\arraybackslash\hspace{0pt}}m{#1}}

\newcommand{\TC}{\text{TRACTOR}}
\newcommand{\embb}{\text{eMBB}}
\newcommand{\mmtc}{\text{mMTC}}
\newcommand{\urllc}{\text{URLLC}}

\definecolor{LightGreen}{RGB}{200,240,200}

\def\BibTeX{{\rm B\kern-.05em{\sc i\kern-.025em b}\kern-.08em
    T\kern-.1667em\lower.7ex\hbox{E}\kern-.125emX}}
    
\begin{document}

\title{From Classification to Optimization: Slicing and Resource Management with \textbf{TRACTOR}\\ 
}

%\IEEEauthorrefmark{1}
\author{
\IEEEauthorblockN{Joshua Groen, Zixian Yang, Divyadharshini Muruganandham, Mauro Belgiovine, Lei Ying, Kaushik Chowdhury} 
}

\maketitle

\begin{abstract}
5G and beyond cellular networks promise remarkable advancements in bandwidth, latency, and connectivity. The Open Radio Access Network (O-RAN) framework offers significant flexibility to reach these competing goals through network slicing and closed-loop RAN control. Central to this evolution is the integration of machine learning (ML) for dynamic network control. This paper presents a comprehensive framework designed to optimize the operation of O-RAN systems. First, we build and share a robust, O-RAN dataset generated from real-world traffic captured across diverse locations and mobility scenarios, replicated within a full-stack srsRAN-based O-RAN system using the Colosseum RF emulator. This dataset serves as a foundation for training and deploying ML models. We then introduce a traffic classification approach that leverages a variety of ML models, showcasing the ability to rapidly train, test, and refine models to improve classification accuracy. With up to 99\% offline average accuracy and 92\% online accuracy for specific slices, the results highlight the flexibility and efficiency of the framework in adapting to different models and network conditions. Finally, we present a physical resource block (PRB) assignment optimization strategy within our framework, which employs reinforcement learning to iteratively refine resource allocation. Our learned policy achieves a higher mean performance score (0.631) than both a manually configured expert policy (0.609) and a random allocation baseline (0.588), demonstrating improved PRB utilization. More importantly, our approach exhibits significantly lower variability, with the Coefficient of Variation (CV) reduced by up to an order of magnitude in three out of four cases, ensuring more consistent and robust performance across varying user configurations. Our contributions, including open-source access to the tools and datasets, enable accelerated research in O-RAN systems and ML-driven network control.

\end{abstract}

% \begin{picture}(0,0)(10,-440)
% \put(0,0){
% \put(0,0){\footnotesize This work has been submitted to the IEEE for possible publication.}
% \put(0,-10){
% \footnotesize Copyright may be transferred without notice, after which this version may no longer be accessible.}}
% \end{picture}

%%% uncomment the following to make the arXiv acceptance notice appear
\begin{picture}(0,0)(10,-520)
\put(0,0){
\put(0,0){\footnotesize This paper has been submitted to IEEE for possible publication.}
\put(0,-10){\footnotesize Copyright may be transferred without notice, after which this version may no longer be accessible.}
}
\end{picture}


\section{Introduction}\label{s: intro}

5G and beyond networks hold the promise of significantly amplified throughput, reduced latency, and a vast increase in connection capacity. However, the pursuit of these objectives often entails competing demands that render a fixed-configuration system incapable of concurrently fulfilling all performance criteria. To address this challenge, 5G introduces the concept of network slicing, comprising three key slices: enhanced Mobile Broadband (eMBB), Ultra Reliable Low Latency Communications (URLLC), and massive Machine Type Communications (mMTC). Network slicing facilitates end-to-end resource allocation tailored to the specific requirements of distinct traffic types. This process involves partitioning a physical network into multiple virtual networks, capitalizing on the growing virtualization trend in 5G networks \cite{8685766}. Each slice is designed with a distinct purpose: eMBB prioritizes maximum bandwidth, URLLC ensures high reliability and minimal latency, while mMTC excels in accommodating a multitude of concurrent connections.

\begin{figure}[t]
    \centering
    \includegraphics[width=\linewidth]{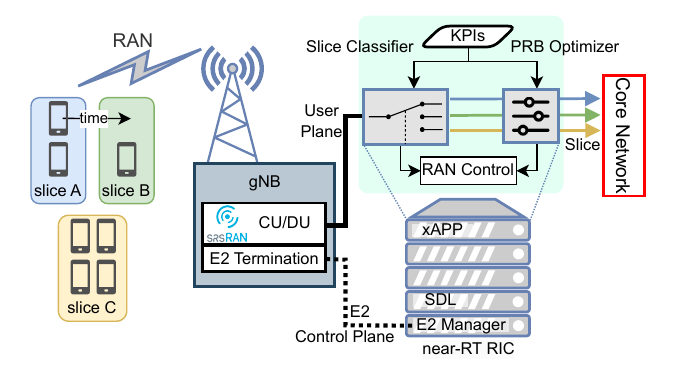}
    \caption{O-RAN system used in $\TC$+ consisting of multiple UEs, a gNB, the near-RT RIC, and an ML based xApp used for traffic classification and PRB optimization. %Each component is implemented in an LXC on top of an SRN within Colosseum. TRACTOR uses O-RAN KPIs created from replayed real user data to enable ML classification of traffic slices. 
    }
    \label{fig:system}
\end{figure}

The Open Radio Access Network (O-RAN) framework supports virtualization as one of its four key architectural principles~\cite{polese2022understanding}. Additionally, O-RAN provides inherent support for closed-loop control driven by machine learning (ML), a capability harnessed through RAN Intelligent Controllers (RICs), shown in Fig.~\ref{fig:system}, for the optimization of network performance. O-RAN introduces standardized technical specifications for open interfaces, enabling the collection of Key Performance Indicators (KPIs) to facilitate diverse control and automation actions. In this context, O-RAN stands out as a robust platform for the implementation of network slicing.

\noindent \textbf{Traffic Classification Challenges:} While significant progress has been made in network slicing (Sec. \ref{sec: related work}), the focus has largely been on improving the RAN component (Fig. \ref{fig:system}) for a fixed number of users with a known traffic type. There are several outstanding issues with current approaches. For example, measurements from deployed 5G networks reveal that the primary throughput and delay bottleneck lies not within the RAN but in the second hop of the network~\cite{narayanan2021variegated,xu2020understanding}—within the LTE Enhanced Packet Core (EPC) or 5G Core Network (CN) outlined in red in Fig.~\ref{fig:system}. Meeting diverse network slice requirements necessitates both end-to-end network path optimization and accurate traffic classification~\cite{QoS-cisco}.

Developing O-RAN compliant traffic slice classification poses several challenges. Neither the traditional 5-tuple flow information (source and destination IP address, source and destination port, and transport layer protocol) nor the complete layer 2 frame are available to the near-real time (near-RT) RIC  in an O-RAN system. Further, the KPIs used in the state-of-the-art 5G classifiers \cite{thantharate2019deepslice, gupta2019machine, khan2022efficient} are neither O-RAN compliant nor based on real 5G traffic.

\noindent \textbf{Resource Allocation Challenges:} Optimizing network slices in near-real time based on the dynamic number of users and their traffic demands presents a significant challenge. 
We propose to use reinforcement learning to deal with this challenge. However,
real-world systems often lack sufficient data during initialization, making it impractical to run exploratory policies solely to gather data before training an optimal policy. This approach is hindered by three primary limitations: (1) the state and action spaces in such systems are vast, making comprehensive exploration infeasible; (2) implementing exploratory policies in live environments risks degraded performance and potential operational costs; and (3) the environment is inherently dynamic, requiring frequent updates to policies. To address these limitations, we employ an iterative pipeline consisting of multiple rounds of training, policy selection, implementation, and data collection to continuously adapt and improve the system.
In each round of this iterative pipeline, we first use multiple algorithms to train on existing data, generating multiple policies. Then we use a policy selection method to select one policy, which will be used for implementation and new data collection for the next round.

We introduce a novel policy selection methodology for each training round. The objective of this approach is to maximize policy performance during implementation in the next round.
Specifically, we divide the existing data into two sets, training set and validation set. The training process uses only the training set, generating multiple policies. 
The validation set will be used to compute the Bellman error for each of these candidate policies, which measures how well the policy has been trained and predicts the policy's performance in the next round.
The proposed method ensures that each selected policy performs well in the next round.

To tackle these challenges, we previously introduced \textbf{TRACTOR}: \textbf{Tr}affic \textbf{A}nalysis and \textbf{C}lassification \textbf{T}ool for \textbf{O}pen \textbf{R}AN \cite{groen2024tractor}. This toolchain leverages O-RAN's open interfaces to collect KPIs from the gNB, enabling near-RT RICs to classify user traffic using ML models while preserving user privacy and system security. TRACTOR is deployed within Colosseum, the world's largest wireless network emulator with hardware-in-the-loop capabilities~\cite{bonati2021colosseum}. 

Building on this foundation, we introduce \textbf{TRACTOR+}, an extended framework that integrates O-RAN E2-compliant control mechanisms to enable dynamic network slicing and resource allocation. Specifically, TRACTOR+ classifies UE traffic in near-real time and automatically reassigns UEs to appropriate slices based on their traffic type. Additionally, we enhance Physical Resource Block (PRB) assignment control, allowing for dynamic PRB allocation across slices to maximize multiple quality-of-service (QoS) objectives and overall network efficiency. This comprehensive approach, implemented within a realistic digital twin environment, enables end-to-end traffic classification, adaptive slicing, and real-time resource optimization for up to 10 UEs.
Our main contributions include the following.

\begin{itemize}[noitemsep,nolistsep]
    \item The creation of a pioneering, publicly available dataset of over 2.83 GB of O-RAN compliant KPIs generated from over 447 minutes of real 5G user traffic \cite{dataset}, along with tools for traffic emulation in Colosseum or similar environments.

    \item An evaluation of an ML-based traffic slice classifier xApp, originally described in \cite{groen2024tractor, belgiovine2024megatron}, demonstrating its feasibility in both offline  (up to 99\% accuracy) and on-line real-time  deployments (up to 92\% accuracy).
    
    % \item Enhancements to the xApp to support multi-UE traffic classification and integrate advanced RB optimization, creating a unified, scalable solution for real-world deployment.
    
    % \item Extensions to the E2 interface and gNB functionalities to enable dynamic UE reallocation between network slices based on real-time traffic classification, along with reinforcement learning-based (RL) PRB optimization to balance per-slice QoS requirements and overall system efficiency.

    \item Enhancements to the xApp and E2 interface to enable multi-UE traffic classification, real-time dynamic UE reallocation between slices, and advanced PRB optimization, providing a unified, scalable solution for adaptive network slicing.

    \item A reinforcement learning-based (RL) PRB optimization framework that iteratively trains, tests, and improves policies using Bellman Error (BE) for policy selection, achieving an increased performance score (0.631) over expert-configured baselines (0.609) and a 66\% reduction in performance variation across a wide range of user combinations.

    \item The development and open-source release of a comprehensive toolkit that enables researchers to efficiently generate custom data sets from O-RAN compliant full-stack emulations, facilitating replication, validation and further research by the community \cite{tractor+_github}.    
\end{itemize}

The remainder of this paper is structured as follows. We discuss the background and related work in Sect.~\ref{background}. We describe our TRACTOR+ tool set and datasets in detail in Sect.~\ref{sec:dataset}. Next, we briefly describe some ML models we deploy to perform traffic classification as an xApp in Sec.~\ref{sec:ml}. Then we describe the RL methods we use for PRB optimization in Sec.~\ref{rl method}.  We present our performance results in Sec.~\ref{sec:results}, followed by a discussion of key lessons learned in Sec.~\ref{discussion}. Finally, we conclude with Sec.~\ref{sec:challenges}.

\section{Background and Related Works}\label{background}

\subsection{O-RAN Principles and Framework}
The near-RT RIC seen in Fig. \ref{fig:system} is a core component of the ML-driven control and optimization of the RAN necessary for 5G. The near-RT RIC hosts xApps, microservices that support custom logic to perform RAN management. Open interfaces are another core principle in O-RAN and one of the key interfaces is the E2 interface that connects the gNB and the near-RT RIC. An xApp can receive KPIs from the gNB over the E2 interface, use those KPIs in a pre-trained ML model, and send back control actions.

\subsection{O-RAN Traffic Classification} \label{sec: related work}
Traditional IP traffic classification relied on packet inspection, as seen in \cite{nguyen2008survey} and \cite{li2021robust}. However, these methods fail when packets are encrypted at the network or data-link layer. To overcome this, statistical traffic properties were used to identify applications, which naturally led to the use of ML in traffic classification, including encrypted traffic. Generally, encrypted traffic classification techniques either use traffic flows defined by a 5-tuple (source IP, source port, destination IP, destination port, and transport-level protocol) \cite{aceto2018mobile} or the entire encrypted packet \cite{lotfollahi2020deep, draper2016characterization, rezaei2019deep} as the input. In an O-RAN system, the near-RT RIC does not have access to the entire packet, encrypted or otherwise. Furthermore, both of these options present a high risk to privacy and security, especially when paired with other information unique to a cellular environment, such as physical location. 

Bonati {\em et al.}~\cite{bonati2021intelligence} select the best performing scheduling policy and resource block assignment in each network slice using deep reinforcement learning fed with data generated in the Colosseum emulator. Both UE assignment and the rewards in the DRL agents are based on knowing the traffic slice \textit{a priori}. Weerasinghe {\em et al.}~\cite{weerasinghe2019supervised} aims to predict incoming traffic at the LTE base station (eNB) using supervised ML. The authors test the performance of their model using simulated bursty LTE traffic data. Li {\em et al.}~\cite{li2017learning} predicts traffic type (among IM, web browsing, and video data) in an upcoming 5 minute period using the previous 3 hours of traffic data in a framework that consists of $\alpha$-stable models, dictionary learning, and alternating direction method (ADM) using 7 million users' OTA 2G-4G application layer data. Johnson {\em et al.}~\cite{johnson2022nexran} perform network slicing and scheduling on an xApp that is based on srsRAN using policy driven heuristic methods.

\noindent {\bf Novelty of Proposed Approach over State-of-the-art: } Our approach to slice allocation differs from user equipment (UE) initiated slicing, which often involves manual or negotiated slice assignments, adding complexity and potential overhead~\cite{choyi2016network, hsieh20215g}. Instead, we focus on network-initiated slicing, allowing the gNB to dynamically allocate traffic flows to the appropriate slice. This approach enhances resource allocation without the need for extensive user involvement. While prior works~\cite{thantharate2019deepslice, gupta2019machine, khan2022efficient} explore similar concepts, none of these utilize O-RAN compliant KPIs generated from real 5G traffic. 

One of the primary shortfalls of previous ML-based slicing is that the KPIs utilized could identify specific users or user traffic, violating user privacy. In contrast, TRACTOR+ does not use any identifying KPIs that could be correlated to a specific UE, source, or destination. This ensures user privacy and reduces the threat surface area for malicious attacks. Further, neither the traditional 5-tuple flow information, the entire layer 2 frame, nor the KPIs used in \cite{thantharate2019deepslice, gupta2019machine, khan2022efficient} are available to the near-RT RIC in an O-RAN system. In contrast to these works, TRACTOR+ is the first work that uses real 5G traces to generate O-RAN compliant KPIs used to perform near-RT traffic slice classification using trained ML models.

\subsection{RL for PRB Assignment}

Optimizing the allocation of Physical Resource Blocks (PRBs) in a dynamic and heterogeneous 5G network environment is critical for meeting diverse Service Level Agreements (SLAs) across slices. Prior works have made significant strides in this domain, with various methodologies focusing on the interplay between resource efficiency and SLA satisfaction. For example, Raftopoulos {\em et al.}~\cite{raftopoulos2024drl} propose a deep reinforcement learning (DRL) agent designed to monitor network performance and adapt slicing policies dynamically. However, their approach is limited to optimizing a single latency-sensitive slice, while our framework simultaneously optimizes PRB allocation across multiple slices, each with distinct SLA requirements.

Similarly, Rezazadeh {\em et al.}~\cite{rezazadeh2023explanation} employ DRL for PRB assignment, using latency as the primary metric across all slices. While effective, this approach relies exclusively on simulated environments and lacks slice-specific performance metrics. In contrast, our method leverages real-world traffic traces and operates within a full-stack RF emulation environment, providing a more realistic and robust evaluation.

Filali {\em et al.}~\cite{filali2023communication} adopt deep Q-learning (DQL) to address RAN resource allocation but focus on assigning PRBs directly to individual UEs rather than optimizing allocations at the slice level. Our work diverges from this by prioritizing slice-level PRB optimization, which better aligns with the principles of network slicing in 5G and beyond.

Several limitations of DRL methods have also been highlighted in the literature. Lotfi {\em et al.}~\cite{lotfi2022evolutionary} note issues such as slow convergence, insufficient exploration, and inefficiencies in dynamic environments. While DRL remains a powerful tool for complex tasks, these limitations underscore the challenges of real-time decision-making in O-RAN systems. Our approach addresses these challenges by integrating slice-specific metrics, enabling rapid adaptation to evolving traffic conditions, and employing an iterative train-test-improve pipeline,
demonstrated in a realistic digital twin environment.

Finally, Bonati {\em et al.}\cite{bonati2021scope} and Polese {\em et al.}\cite{polese2022colo} utilize DRL for PRB allocation but do so with fixed traffic profiles and static user distributions. These constraints limit applicability in real-world scenarios, where traffic patterns and user demands are inherently dynamic. In contrast, our method dynamically adjusts PRB allocation based on real-time metrics and varying user loads, ensuring optimal performance across all slices.

As mentioned in Section~\ref{s: intro}, one key challenge of using RL in PRB assignment is the lack of sufficient data covering different user loads. In the literature, people either use some exploratory policy to collect a large amount of data before training or train with online exploration.
Both methods risk degraded performance and potential operational costs.
We propose an iterative train-test-improve pipeline that consists of multiple rounds of training, policy selection, implementation and data collection to deal with this challenge, which ensures good and gradually improving performance in each round.

\section{Creating an O-RAN Dataset}\label{sec:dataset}
To understand real 5G traffic patterns and overcome one of the challenges found in prior work, we create a new and publicly available tool chain and dataset. We implement a software-defined RAN using the srsRAN-based SCOPE framework \cite{bonati2021scope} for both the gNB and one or more UEs, as depicted in Fig. \ref{fig:system}. SCOPE extends srsLTE (now srsRAN) version 20.04 by introducing features like an E2 interface and open APIs for real-time gNB reconfiguration, along with additional data collection tools. Our near-RT RIC, part of the ColO-RAN framework \cite{polese2022colo}, utilizes a minimal near-RT RIC based on the O-RAN Software Community's near-RT RIC (Bronze release). This near-RT RIC is structured as Docker containers within the ColO-RAN LXC, providing essential functionalities such as E2 interface support for data collection and control communication with RAN nodes, alongside a sample xApp for collecting fundamental KPIs from the gNB.

We expand the existing framework by significantly modifying both the gNB and the E2 interface to enable more flexible and dynamic network management. Specifically, we update the gNB code to allow seamless reassignment of UEs between network slices. Additionally, we enhance PRB allocation, which was previously fixed at gNB startup, by enabling dynamic PRB assignment during runtime. These improvements support our goal of adaptive, network-initiated slicing to optimize resource allocation efficiency.

To complement these modifications, we extend the E2 message format and parsing capabilities to include new control fields for directing slice assignments of individual UEs and updating slice PRB allocations. These updates empower the near-RT RIC to dynamically manage UE allocations across slices with greater precision, improving its ability to meet varying SLA requirements.

Beyond these structural changes, we develop a custom traffic generator and the TRACTOR+ xApp framework (highlighted in light green in Fig.~\ref{fig:system}), along with various supporting utilities, forming part of the publicly available TRACTOR+ code base~\cite{tractor+_github}. This comprehensive tool set provides researchers with an open framework for exploring traffic classification, advanced slicing strategies, and dynamic resource allocation algorithms.

\subsection{Collecting Real-World 5G User Data}\label{sec:real world 5G}

\begin{figure}[htb]
    \centering
    \includegraphics[width=0.6\linewidth]{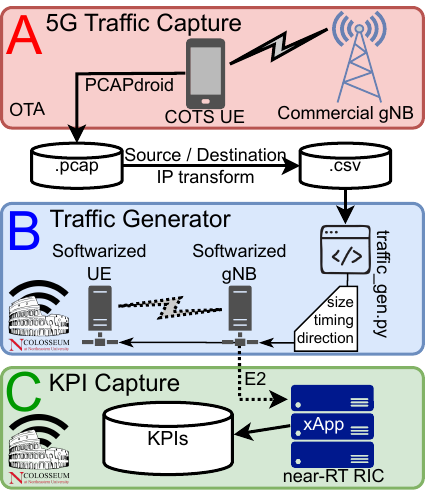}
    \caption{Process to create an O-RAN compliant dataset.}
    \label{fig:create dataset}
\end{figure}

To collect real-world 5G user traffic, we use the open source PCAPdroid \cite{pcapdroid} Android application on a Google Pixel 6 Pro smartphone and generate packet captures (.pcap files) of user traffic. This is illustrated by block A in Fig. \ref{fig:create dataset}. We use a variety of applications for each network slice. For eMBB, we stream videos, browse the Internet, and transfer large files. For URLLC, we conduct both voice phone calls, video chat, and utilize real time AR applications. For mMTC we capture texts and background traffic from all apps when the phone is not actively being used. This is not the typical example of mMTC traffic, such as IoT applications. However, it does fit nicely in the fundamental definition of mMTC because it is low throughput, latency tolerant communication from numerous applications. PCAPdroid provides a custom trailer that adds metadata identifying the phone application to each packet capture. This data is used to ensure captured data is labeled with the correct network slice. %It is possible to use the \texttt{pcapdroid.lua} plug-in for Wireshark to view the application used in the pcap capture. 
This large dataset was collected on multiple different days, in different locations, with different levels of mobility. Table \ref{tab:datacap} gives a detailed overview of the parameters used to capture 447 minutes of 5G traffic.

\begin{table}[hb]
    \centering
    \resizebox{\linewidth}{!}{%
    \begin{tabular}{|l;{1pt/1pt}l;{1pt/1pt}l;{1pt/1pt}l;{1pt/1pt}l|} \hline
    Slice & Application & Location & Mobility & Time (min) \\ \hline\hline
    eMBB & Chrome, YoutTube, One Drive & Residential & Stationary & 43.5 \\
    eMBB & YouTube & Campus & Stationary & 29.0 \\
    eMBB & YouTube & Campus & Stationary & 17.2 \\
    eMBB & Netflix & Campus & Stationary & 21.3 \\
    eMBB & One Drive & Campus & Stationary & 30.6 \\
    eMBB & YouTube & Campus & Stationary & 4.9 \\
    eMBB & Pandora & Campus & Stationary & 6.7 \\
    eMBB & One Drive & Campus & Stationary & 1.1 \\
    eMBB & Chrome & Campus & Stationary & 5.7 \\ \hline\hline
    mMTC & background & Mixed & Driving & 64.0 \\
    mMTC & background & Campus & Walking & 11.8 \\
    mMTC & background & Campus & Stationary & 23.7 \\
    mMTC & background & Campus & Stationary & 23.9 \\
    mMTC & background & Campus & Stationary & 16.4 \\
    mMTC & background & Campus & Stationary & 5.6 \\
    mMTC & background & Campus & Stationary & 20.9 \\ \hline\hline
    URLLC & Google Meet & Campus & Walking & 57.0 \\
    URLLC & Phone, Google Meet & Residential & Walking & 5.8\\
    URLLC & Google Meet & Campus & Stationary & 8.0 \\
    URLLC & Facebook Messenger & Campus & Stationary & 21.0 \\
    URLLC & Google Meet & Campus & Walking & 7.9 \\
    URLLC & Google Meet & Campus & Stationary & 7.1 \\
    URLLC & Google Maps Live View AR & Campus B & Walking & 6.5 \\
    URLLC & Facebook Messenger & Campus & Stationary & 3.9 \\
    URLLC & Microsoft Teams & Campus & Stationary & 3.5 \\ \hline
    \end{tabular}
    }
    \caption{Detailed break down of real world data capture variables including application used, location, and mobility. For a given traffic slice, each row was collected on a different day.}
    \label{tab:datacap}
\end{table}

\subsection{Replaying Traffic in Colosseum}
After capturing real-world data, we need a method to replay traffic in an O-RAN system to capture KPIs. To enable this, we use Wireshark to export the packet dissection to a .csv file. Next, we alter the source and destination IP addresses so that IP traffic is between two endpoints. This is especially necessary for the mMTC traffic where, in reality, the UE is reaching out to a wide range of IPs. These transformation processes are shown between blocks A and B in Fig. \ref{fig:create dataset}. While we chose to use real traces from an UE using deployed 5G networks, any .pcap capture can be used as the input to the system. In this way, we enable researchers to use any network packet capture as the input to understand the impact to O-RAN KPIs.

We built a traffic generator tool to replay the traffic between the UE and gNB as shown in block B in Fig. \ref{fig:create dataset}. The traffic generator emulates the original traffic by reading the length field of the .csv file and sending a random byte string of the appropriate length at the time indicated by the packet timestamp. We use User Datagram Protocol (UDP) in our traffic emulation because any transport layer functions are already captured in the original traffic. In order to ensure the replay traffic is the same as the original, we subtract 70 bytes of overhead to remove the PCAPdroid trailer and account for the new Ethernet frame, IP header, and UDP header we use. We use the source and destination IP to determine if the UE or gNB should be sending the traffic. We utilize the time stamp to wait the appropriate amount of time before sending the next packet. In this way, the UE and gNB accurately emulate the first hop (cellular RAN) of the original 5G traffic. 

We have developed several supplementary utilities related to traffic generation, including multiple UE traffic generation, anomalous traffic generation, and the introduction of arbitrary RF interference. The multiple UE traffic generator operates by replaying a randomly selected trace for each UE participating in the experiment. While our original traces already encompass real network competition, this approach enables researchers to delve deeper into the influence of multiple competing UEs on O-RAN KPIs. The anomalous traffic generation tool offers support for two potential attack scenarios described in detail in \cite{rumesh2024federated}. Firstly, it models a DoS UDP flood attack using a statistical model. Secondly, it facilitates an attack known as \say{data-hog,} which employs the original traces but increases packet sizes based on a Gaussian distribution. The RF interference tool generates arbitrary waveforms specified by the user on the uplink or downlink channel. This provides a valuable tool for understanding the impact on O-RAN systems as the RF environment becomes ever more congested. These tools serve as starting points for exploring security implications within O-RAN systems.

We also enable multiple UEs in different slices replaying different slice specific traffic traces simultaneously. Our traffic generator can randomly select one of the traces from a given slice and assign that trace to a specific UE. This system supports up to 10 different UEs per gNB. Each of the three network slices we configure supports from 0 to 10 UEs. This provides 285 unique combinations of number of UEs in each slice, with at least 1 UE total. When combined with number of available traces in each slice, this creates an incredibly diverse, realistic emulation environment used to train, test, and validate ML models.

\subsection{Capturing KPIs}\label{capturing kpis}
The traffic generation script allows us to replicate the timing, length, and direction of all data sent between the UE and gNB, while completely anonymizing the actual payload within our experimental test bed. Our O-RAN test bed further emulates the channel conditions between the gNB and UE based on measured channel conditions for a real deployed cellular system. This allows us to accurately capture the O-RAN KPIs as if the original communication were taking place in our O-RAN test bed. To the best of our knowledge, this is the first dataset generated with live 5G traffic and accurately replayed in an O-RAN system to capture KPIs.  

To capture KPIs, we employ the TRACTOR+ xApp framework, which retrieves requested KPIs from the gNB every $250\mathrm{ms}$ over the E2 interface. This xApp uses our ML model for online traffic slice classification. Simultaneously, we record all the available KPIs for offline training. These KPIs are stored in a .csv file and are part of our publicly accessible dataset. The process is depicted in block C of Fig. \ref{fig:create dataset}. Additionally, we provide a tool for automated IPsec configuration over this E2 interface, as elaborated in \cite{groenCost}, which is the first open-source solution for configuring O-RAN compliant IPsec over the E2 interface, facilitating swift O-RAN system deployment and systematic performance analysis.

\subsection{Pre-processing KPIs} 
In our O-RAN setup, we have access to 31 KPIs (listed in~\cite{tractor+_github}) that encompass various low-level performance metrics, including identifiers such as IMSI, RNTI, and slice ID. Before feeding these KPIs into our ML model, we pre-process the data to remove uniquely identifying information and certain administrative fields, such as slice assignments and scheduling policies, to preserve user privacy and confidentiality. Additionally, we exclude KPIs like the received signal strength indicator (RSSI), which lack meaningful values in our Colosseum emulation, thereby reducing input dimensions without loss of relevant information. The final dataset used for model training consists of 17 carefully selected KPIs, as detailed in Table \ref{tab:kpi}. A full list of KPIs and their descriptions is available in our public dataset.

%The second major step is to trim some of the time from the beginning and the end of the KPI capture. While the replay script exactly replicates the original capture, there are periods of time before and after the replay script runs when KPIs are still being captured. Further, within some slices there can be large periods of no traffic. We had to manually inspect the KPIs and remove these periods of silence for training purposes only. 

\begin{table}[hb]
\resizebox{\columnwidth}{!}{
    \centering
    \begin{tabular}{| l | l |}
        \hline 
        \textbf{KPI name} & \textbf{Description} \\
        \hline \hline
        dl\_mcs & Downlink modulation and coding \\
        \hline
        dl\_n\_samples & Number of download samples in previous 250 ms \\
        \hline
        dl\_buffer\_bytes & Downlink queue length in bytes \\
        \hline
        tx\_brate\_downlink\_Mbps & Downlink bitrate in Mbps \\
        \hline
        tx\_pkts\_downlink & Downlink number of packets transmitted in previous 250 ms \\
        \hline
        dl\_cqi & Downlink channel quality indicator \\
        \hline
        ul\_mcs & Uplink modulation and coding \\
        \hline
        ul\_n\_samples & Uplink number of samples in previous 250 ms \\
        \hline
        ul\_buffer\_bytes & Uplink queue lengith in bytes \\
        \hline
        rx\_brate\_uplink\_Mbps & Uplink bitrate in Mbps \\
        \hline
        rx\_pkts\_uplink & Uplink number of packets recieved in previous 250 ms \\
        \hline
        rx\_errors\_up\_perc & Uplink percent of packets with errors in previous 250 ms \\
        \hline
        ul\_sinr & Uplink signal to interference and noise ratio \\
        \hline
        phr & UE power head room \\
        \hline
        sum\_reqsted\_prbs & Sum of the resource blocks requested in previous 250 ms \\
        \hline
        sum\_granted\_prbs & Sum of the resource blocks granted in previous 250 ms \\
        \hline
        ul\_turbo\_iters & Uplink turbo encoding \\
        \hline
        
    \end{tabular}
    }
    \caption{$\TC$ uses 17 O-RAN compliant KPIs. None of these KPIs expose uniquely identifiable information.}
    \label{tab:kpi}

\end{table}

Our TRACTOR+ xApp framework maintains a structured dictionary that stores a sliding window of KPI values alongside metadata such as timestamps, each UE’s currently assigned slice, and the number of PRBs assigned to each slice. Since srsRAN supports variable channel bandwidths—where the total number of available PRBs depends on the specific bandwidth—it uses a fixed-size bitmap to define PRB allocation per slice. To ensure consistent control across different bandwidths, our framework adopts this bitmap representation, referred to as resource bits (rB), which map to LTE PRBs. This structure enables seamless adaptation across varying network conditions, with adjustments handled through a scaling factor.

Beyond maintaining structured KPI data, our framework provides a simple API for diverse ML research applications. The stored KPIs enable per-UE traffic classification, while the slice assignment metadata allows us to track UE distribution across slices for PRB optimization. Additionally, by analyzing memory usage and inference times as a function of the number of KPIs and UEs included in an experiment, we can systematically benchmark performance across different configurations. This flexibility makes TRACTOR+ a valuable tool for evaluating ML-driven network control strategies.

\subsection{Train-Test-Improve}\label{ss:tti}

As discussed earlier, optimizing network slices in near-real time is complicated by the vast state and action spaces, the operational risks of running exploratory policies in live environments, and the dynamic nature of real-world systems. These challenges make it impractical to rely on one-time training or static policies. Instead, we adopt an iterative approach to continuously refine performance. This approach combines training, policy selection, real-world testing, and iterative improvements, ensuring the system adapts effectively to evolving traffic demands and user behavior.

Throughout this process, we use a tuple to represent how many users are in each slice, e.g: (\#mMTC, \#URLLC, \#eMBB). Because the total number of possible combinations (or user tuples) is very large, we started with a smaller subset of more common user tuples. This set of nine user tuples consists of: (0, 1, 2), (0, 2, 2), (1, 1, 2), (1, 1, 4), (1, 2, 1), (1, 2, 3), (1, 2, 5), (1, 3, 4), (3, 2, 3) and is used in every round of training and testing. Additionally, in each round of training there are other user tuples that are introduced, representing less common combinations of users.

In each round, or epoch, we start with one or more PRB allocation policies. For each UE in the tuple we use the TRACTOR+ framework to re-create the appropriate traffic type. Once all the UEs begin their traffic flow, we record the KPIs generated at the gNB for 2 minutes. Because many of the traffic traces are significantly longer than 2 minutes, we break the longer traces up into 2 minute sub-trace chunks. This further increases the variety in specific traffic replayed, ensuring each 2 minute trial represents a realistic but random snapshot in time. For each round of experiment, we collect one or more trials for each user tuple. In this way we create a very large, publicly available dataset~\cite{dataset} of over 2.83 GB, consisting of 4,000 different trials.

% After we collect the data for one epoch, we calculate the current performance of the algorithms under test. Specifically, we evaluate the performance based on slice-specific KPIs such as latency, throughput, and resource efficiency, ensuring that the metrics align with the SLA requirements of each slice. Once the evaluation is complete, the dataset is used to re-train a new set of PRB allocation policies using RL algorithms. 
% Then we select a policy based on the metric of Bellman error and the selected policy will be used for implementation and new data collection for the next round.
% By incorporating both successful and underperforming trials, we allow the learning process to refine its decision-making and adapt to real-world traffic conditions more effectively. Complete details on calculating performance, training new policies, and policy selection are presented in Section \ref{rl method}.

After collecting data for one epoch, we evaluate the performance of the tested algorithms based on slice-specific KPIs such as latency, throughput, and resource efficiency, ensuring that the metrics align with the SLA requirements of each slice. The collected dataset is then used to re-train a new set of PRB allocation policies using RL algorithms.  
Once training is complete, we select the optimal policy based on the Bellman error metric, which quantifies the expected improvement in decision-making. The selected policy is then deployed for the next round of data collection, allowing the system to iteratively refine its PRB allocation strategy.  
By incorporating both successful and under-performing trials, the learning process continuously improves, adapting to real-world traffic conditions more effectively. Complete details on performance evaluation, policy training, and selection are provided in Section~\ref{rl method}.

To streamline this iterative pipeline, we developed several automation scripts that integrate dataset generation, KPI logging, and policy deployment into a single workflow. These scripts significantly reduce overhead by enabling rapid deployment, testing, and continuous improvement of PRB allocation algorithms. Additionally, the automation ensures consistency across experiments, minimizes human intervention, and facilitates reproducibility for future research. All our scripts, along with the TRACTOR+ framework, are open-sourced to encourage collaboration and accelerate advancements in traffic classification and resource allocation.

This iterative process of training, testing, and improving ensures that our approach remains robust in dynamic, real-world environments. By systematically refining the allocation policies over multiple rounds, the system can achieve better performance across all slices while maintaining flexibility to adapt to unexpected traffic patterns. This continuous improvement cycle demonstrates the scalability and practicality of our framework for near-real-time resource management in O-RAN systems.

\section{ML for Traffic Classification}\label{sec:ml}

\begin{figure}[htb]
    \centering
    \includegraphics[width=\linewidth]{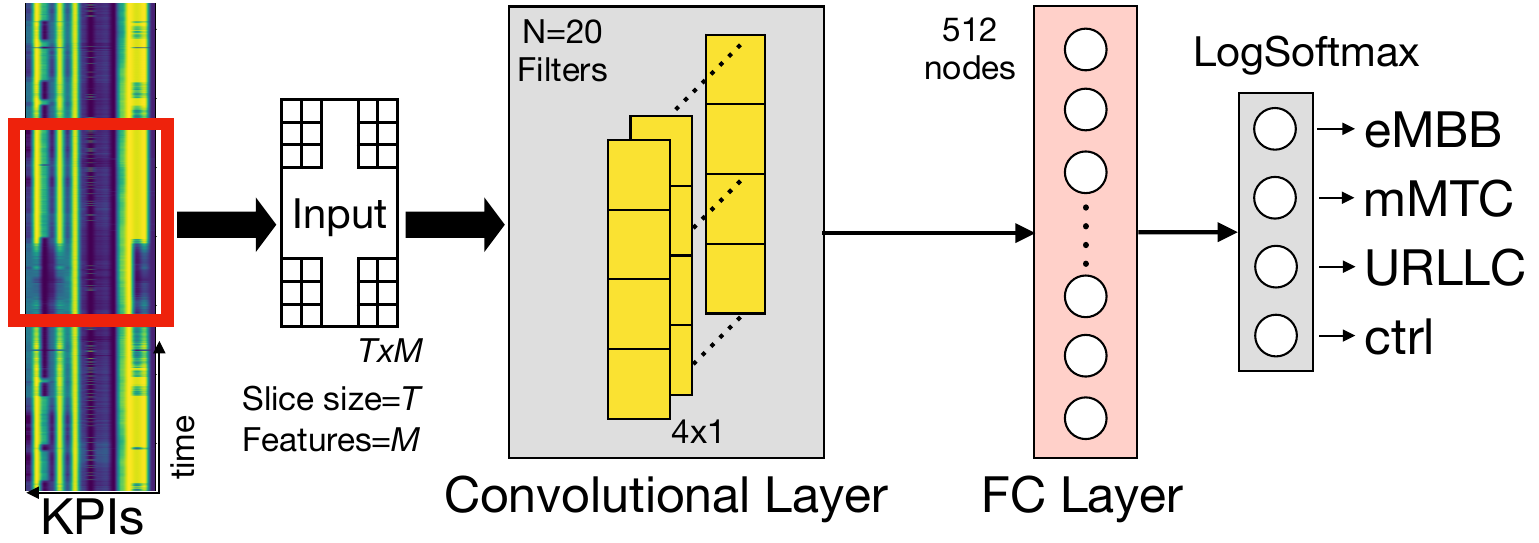}
    \caption{CNN model used for traffic classification in the TRACTOR framework.}
    \label{fig:ICC_arch}
\end{figure}

To demonstrate the the potential of the TRACTOR+ tool-set, we train and deploy an xApp that uses multiple ML models for traffic classification; a CNN shown in Fig.~\ref{fig:ICC_arch} and a Transformer model described in \cite{belgiovine2024megatron}, to perform traffic slice classification. 

% \begin{figure}[htb]
% \centering
%   \includegraphics[width=\linewidth]{img/intro_fig.jpg}
%     \caption {MEGATRON~\cite{belgiovine2024megatron}, a transformer-based 5G traffic classification system used in the TRACTOR framework.}
%   \label{fig:transformer}
% \end{figure}

\subsection{Data Pre-processing}
We form the input to the traffic classification model by stacking $T$ consecutive KPIs sampled by the gNB, highlighted on the left side of Fig.~\ref{fig:ICC_arch}. Each KPI feature set is formed by $M=17$ traffic metrics listed in Table ~\ref{tab:kpi}. Thus, when stacked, we obtain a $T \times M$ 2D input, representing a snapshot of how such traffic indicators evolve over a $T \times 250\,\mathrm{ms}$ time span. In other words, we use a sliding window with the size of $T \times 250\,\mathrm{ms}$, making a new prediction at every $250\,\mathrm{ms}$. We collect traffic metrics in Colosseum for the three main traffic categories, i.e. $\embb$, $\urllc$, and $\mmtc$, and include an additional class of samples, denoted as \textit{control} (ctrl), or ``silent", class to identify the portions of traffic where no application data is being exchanged between registered UEs and gNB. Note that depending on the UE's activities, control class traffic might be found in any of the three traffic categories and therefore we consider \textit{ctrl} as a fourth meta-class that can be used to identify idle users and applications. Adding this traffic slice may benefit users and service providers by providing a separate slice with minimum resources allocated to keep connections alive and preserve resources for the other slice types. 

We pre-process input slices in the dataset by normalizing each KPI feature individually, in order to keep all values in the dataset within $[0,1]$ interval. We then randomize the order of slices and allocate $80\%$ of input samples to train the model, while retaining $20\%$ for testing purposes that the model does not see during training. After training and testing datasets are defined, we obtain $111.6$K training samples and $27.9$K testing samples, equally distributed among the main 3 traffic classes and control traffic class.

\subsection{Model Training}
To train our models, we slice the KPI data into groups of $T$ consecutive time samples for each KPI. Both the CNN and transformer-based models are designed to classify these slices of KPI traffic patterns, though their architectures and approaches differ.

The proposed CNN model begins with a single 2D convolutional layer containing $20$ kernels of size $4 \times 1$ to learn local patterns at each individual feature value, as the KPI features are not spatially correlated. This is followed by one fully connected (FC) hidden layer of $512$ neurons using ReLU activation, and a LogSoftmax output layer for classification. We train the CNN model using the Adam optimizer, starting with an initial learning rate of $10^{-3}$, which is reduced by a factor of $10$ upon reaching a plateau in cross-entropy loss, with a minimum learning rate of $10^{-5}$. %Training is performed for up to $350$ epochs with an early stopping criterion to prevent overfitting.

We also trained a transformer model detailed in~\cite{belgiovine2024megatron}, that processes the same $T$ consecutive time samples of KPIs as a sequence, treating the KPIs sampled at each time step as a \textit{token} within the sequence. These sequences are passed through transformer encoder layers, which generate contextualized representations for each token. The output of the encoder layers is flattened and fed into a fully connected layer with $256$ neurons using ReLU activation, followed by a LogSoftmax output layer to compute the log probabilities for traffic classification. Like the CNN, the transformer model is trained using the Adam optimizer with an initial learning rate of $10^{-2}$, reduced upon encountering a plateau in loss. %Early stopping is applied here as well, ensuring efficient training.

Both models are trained for up to 350 epochs, with early stopping to prevent overfitting. All computations run on an NVIDIA A100 GPU. The transformer model achieves inference times under 1 ms on a GPU and 1.5 ms on a CPU per user. In contrast, the CNN, when processing multiple UEs in the full TRACTOR+ xApp, averages 2.27 ms. Since KPIs are sampled every 250 ms, both models operate within real-time constraints, adding no latency. This ensures they meet xApp requirements, where event reaction times range from 10 ms to 1 s.

\section{RL for PRB Optimization}\label{rl method}
Reinforcement Learning (RL) has emerged as a promising solution for optimizing Physical Resource Block (PRB) allocation in 5G networks. Its ability to adapt to dynamic and heterogeneous environments makes it particularly well-suited for managing the complexities of network slicing, where diverse Service Level Agreements (SLAs) must be simultaneously satisfied across slices. Unlike static or heuristic-based methods, RL learns policies that can leverage real-time feedback to optimize resource allocation under varying traffic conditions, enabling a more efficient and adaptive approach.

Prior research demonstrates the potential of RL for PRB optimization, focusing on key challenges such as balancing resource efficiency with SLA satisfaction and adapting to dynamic network demands. However, existing methods often fall short when applied to real-world scenarios due to limitations such as reliance on simulated environments, single-slice optimization, or static traffic assumptions. Building on these works, our framework addresses these gaps by incorporating slice-specific performance metrics and leveraging real-world traffic traces in a realistic RF emulation environment. This ensures robust, real-time adaptation and optimization, paving the way for a more practical deployment of RL in O-RAN systems.

\subsection{Performance Metrics}

To design an effective observation space for the RL agent, we consider two critical factors. First, the agent must observe relevant KPIs that are strongly correlated with the performance objectives of each slice, such as end-to-end latency, throughput, and reliability. Second, the dimensionality of the observation space must remain manageable, as a high-dimensional input can degrade the performance of ML-driven xApps by increasing computational complexity and leading to suboptimal policy learning~\cite{raftopoulos2024drl}.

In this paper, each network slice is tailored to meet a distinct set of performance objectives. The eMBB slice prioritizes maximum bandwidth to accommodate high-throughput applications such as video streaming or file downloads. In contrast, the URLLC slice focuses on ensuring low latency to support real time communication applications. Finally, the mMTC slice is designed to handle a large number of low-bandwidth, latency-tolerant connections, making it suitable for IoT deployments. These distinct objectives guided our selection of KPIs and the creation of unique performance functions for each slice.

\textbf{eMBB:} The primary goal of eMBB is to maximize throughput. Unlike URLLC, this slice is more tolerant to latency, allowing for some packet queuing. Additionally, when allocating PRBs it is important to avoid over-provisioning beyond actual traffic demands. To address these considerations, we designed a performance function based on the change in the down-link buffer length during each observation period. A neutral score corresponds to maintaining the same queue length, while draining the queue yields a higher score and adding to the queue results in a lower score.

We estimate the change in queue length using the transmission rate ($tx\_brate$, in Mbps) and the down-link buffer size ($dl\_buffer$, in bytes) from the previous observation period as shown in Equation~\ref{eqn:embb}. 
\begin{equation}\label{eqn:embb}
  \text{score} = \alpha \Big( \beta + (tx\_brate \times T) - (dl\_buffer \times \kappa) \Big)
\end{equation}
The score for eMBB is measured in bits. Here, $T$ is the measurement period, and $\kappa$ is a conversion factor.
This score represents the net number of bits added to or drained from the queue over the prior observation period $T$. To ensure the score falls within the range $[0, 1]$, we apply a scaling factor $\alpha$ %= \frac{1}{3}$ 
and an offset $\beta$ %= \frac{3}{2}$
. %These values were determined empirically based on prior datasets, reflecting typical downlink throughput and buffer sizes. 
Finally, any values outside $[0, 1]$ are clipped to ensure consistency across all performance metrics.

\textbf{URLLC:} The broad goal of URLLC is to minimize end-to-end latency. However, the near-RT RIC does not have direct access to end-to-end latency measurements, as these are application-layer KPIs available only to the end points through the core network \cite{raftopoulos2024drl}. Consequently, xApps cannot obtain such measurements from the gNB directly. Despite this limitation, the near-RT RIC can utilize proxy metrics that approximate end-to-end latency. For instance, by calculating the average time a bit spends in the transmission queue, we can estimate the latency introduced by the RAN. This can be derived using the down-link buffer size and the transmission rate. Specifically, the time spent in the queue is the buffer length divided by the transmission rate. 
We use this to define a URLLC performance function in Equation~\ref{eqn:urllc}. 
\begin{equation}\label{eqn:urllc}
    \text{score} = \frac{1}{\gamma} \max\left(0, \gamma - \frac{dl\_buffer (\text{bytes}) \times \kappa}{tx\_brate (\text{Mbps})}\right)
\end{equation}
%where $\gamma = 1$ in our experiments. 
This metric (with units of seconds) yields a score of 1 when there is no queuing delay in the down-link buffer and decreases linearly (down to 0) as queuing delay increases. Notably, $\gamma$, which represents the maximum allowable delay in seconds, can be adjusted to reflect different delay tolerances while still ensuring the score remains within $[0, 1]$. 
In rare cases where both the down-link buffer and transmission rate are zero, we assign a score of 1. Although infrequent in our URLLC experiments, such a scenario implies no RAN-induced latency and therefore should not be penalized.

\textbf{mMTC:} The goal of mMTC is to support as many users as possible, with each user having low bandwidth requirements and relatively high tolerance for latency. This means that efficient resource utilization is prioritized over throughput or latency performance. To achieve this, we use the ratio of PRBs granted to PRBs requested as the primary performance metric, reflecting how effectively the slice meets demand.

However, mMTC traffic patterns are highly bursty, characterized by long periods of inactivity followed by sudden surges in demand. To prevent over-provisioning during idle periods, we introduced an additional check. If no PRBs are requested ($prb\_req = 0$), we assign a score based on the current resource allocation: 
$\text{score} = \frac{1}{slice\_prb}$, 
where $slice\_prb$ represents the number of PRBs currently allocated to the mMTC slice. This encourages minimizing resource allocation when the slice is idle.

When PRBs are requested ($prb\_req > 0$), the score is calculated as:
$\text{score} = \min(1, \frac{prb\_granted}{prb\_req})$,
ensuring the utilization score remains within $[0, 1]$. A score of 1 indicates that all requested PRBs were granted, while lower scores reflect under-provisioning.

Combining these, we get the performance score for mMTC, shown in Equation~\ref{eqn:mmtc}, designed to maintain sufficient PRBs for active UEs while avoiding excessive resource allocation during idle periods.
\begin{equation}\label{eqn:mmtc}
    \text{score} = 
\begin{cases} 
\frac{1}{slice\_prb} & \text{if } prb\_req = 0 \\
\min\left(1, \frac{prb\_granted}{prb\_req}\right) & \text{if } prb\_req > 0
\end{cases}
\end{equation}
This approach balances resource efficiency with the need to handle bursts of traffic effectively, ensuring the mMTC slice is neither over- nor under-provisioned.

\subsection{RL Methods}\label{ss:RL_methods}

\textbf{Model for RL:} To optimize PRB allocation using RL algorithms, we use the following Markov Decision Process (MDP) to model the PRB allocation problem. We define one period as $T = 250$ ms. We define the state $s$ in each period by
\begin{align*}
    s & = (\text{\# mMTC users}, \text{\# URLLC users}, \text{\# eMBB users},\nonumber\\
    & \text{\# mMTC RBs}, 
    \text{\# URLLC RBs})
\end{align*}
where one resource bit (Rb) maps to three PRBs except that for eMBB, the last Rb contains only two PRBs. 
We do not include the number of Rbs in eMBB in the state because we know that the total number of Rbs is $17$ and
\[
\text{\# Rbs in eMBB} = 17 - \text{\# Rbs in mMTC} - \text{\# Rbs in URLLC}.
\]
In each period, based on the state $s$, we define the action $a$ in Table~\ref{tab:action}.

\begin{table}[htb]
\centering
\rowcolors{2}{gray!20}{}
\begin{tabular}{|l;{1pt/1pt}l|}
\hline
Action $a$ & Meaning  \\
\hline\hline
0 & Keep the current RB allocation  \\
1 & Move one RB from mMTC to URLLC  \\
2 & Move one RB from mMTC to eMBB  \\
3 & Move one RB from URLLC to mMTC \\
4 & Move one RB from URLLC to eMBB  \\
5 & Move one RB from eMBB to mMTC   \\
6 & Move one RB from eMBB to URLLC \\
\hline
\end{tabular}
    \caption{Definition of the action $a$.}
    \label{tab:action}
\end{table}

The state transition is deterministic and the next state $s'$, that is the user tuple and Rb allocation in the next period, will be determined by the current state $s$ and current action $a$. Recall that for each slice we define a score function in the previous subsection. The reward $r$ is defined as the average score over the three slices in the next period. The discount factor $\gamma$ is $0.99$.

\textbf{RL Algorithms:} We use the tabular Q-learning algorithm~\cite{watkins1989learning} and the Deep Q-network (DQN) algorithm~\cite{mnih2015human}. For the tabular Q-learning algorithm, we use a learning rate $0.1$ for each training step. 
For DQN algorithm, we normalize the state by dividing the number of users by $10$ (the largest number of users that the system supports) and dividing the number of Rbs by $17$ (total number of Rbs). We use two-layer fully-connected neural networks with size $5\times 256\times 7$.
We use the Adam~\cite{kingma2014adam} optimizer with learning rate $0.01$. Note that we can also use other RL algorithms such as actor-critic-type algorithms. In this paper, we focus on these two RL algorithms to demonstrate the effectiveness of the proposed train-test-improve pipeline, which will be detailed in the following paragraph.

\begin{figure}[htb]
    \centering
    \includegraphics[width=1.0\linewidth]{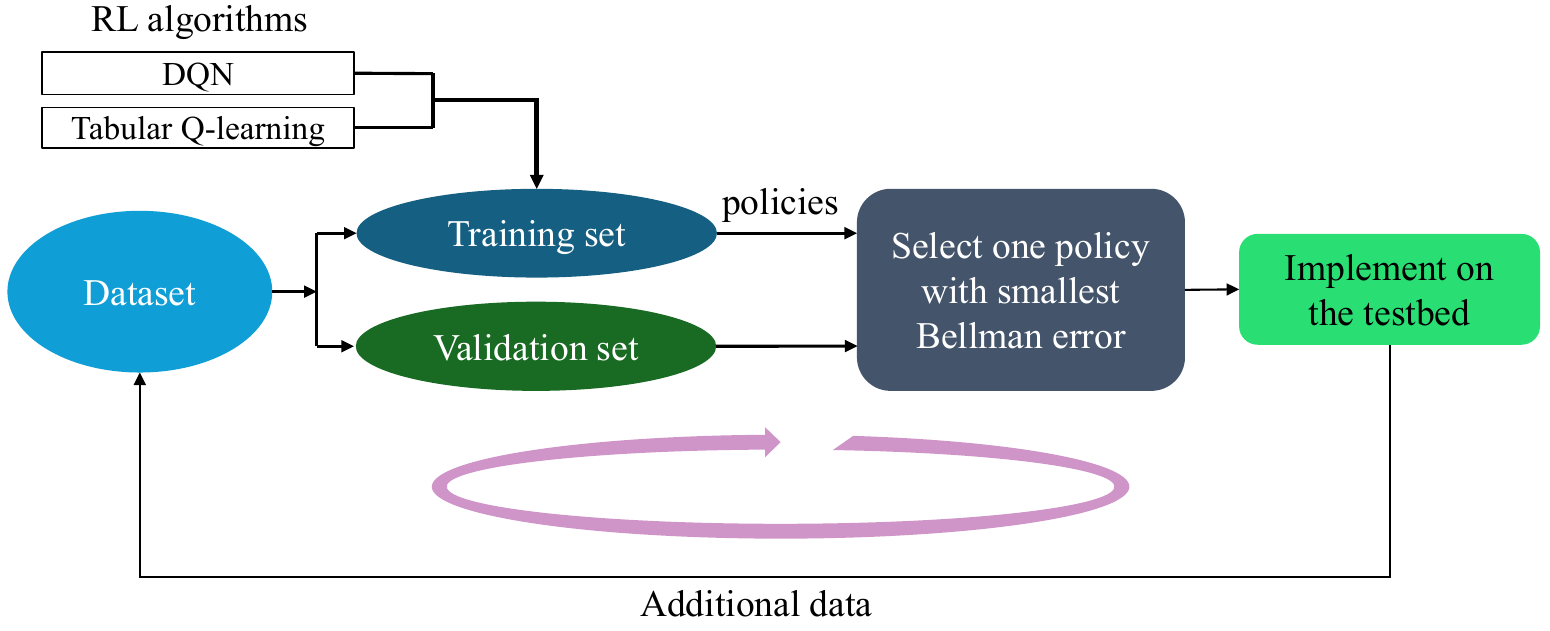}
    \caption{Illustration of the train-test-improve pipeline, an iterative RL-based learning process. At each epoch, new policies are trained, evaluated using Bellman error, and the best-performing policy is deployed for data collection in the next round, continuously refining performance.}
    \label{fig:train-test-improve}
\end{figure}

\textbf{Train-Test-Improve Pipeline:} As shown in Figure~\ref{fig:train-test-improve}, the train-test-improve pipeline is an iterative learning process using RL algorithms. Suppose initially we have a dataset with limited amount of data, which can be generated using some initial policy. We call it round/epoch 1. We first divide the dataset into two sets, a training set and a validation set. Using RL algorithms to train on the data in the training set, we can obtain one policy for each RL algorithm. In this paper, we use the tabular Q-learning algorithm and the DQN algorithm, generating two policies. Next, we compute the Bellman error for each policy using the data in the validation set. Note that for each policy, there is an estimate of the optimal Q-function, denoted by $\hat{Q}(s,a)$.
The (average) Bellman error (BE) is defined as
\begin{align*}
    \text{BE} = \frac{1}{|{\cal D}_{\mathrm{val}}|}\sum_{(s,a,s',r)\in {\cal D}_{\mathrm{val}}} \left| \hat{Q}(s,a) -  (r+ \gamma \max_{a'} \hat{Q}(s',a') ) \right|,
\end{align*}
where ${\cal D}_{\mathrm{val}}$ is the validation set. Then, we select the policy with the smaller BE to implement on the testbed.
A smaller BE indicates a lower estimation error of the Q-function. Since the true optimal Q-function corresponds to an optimal policy,
a more accurate Q-function generally leads to better performance.
Thus, the selected policy is expected to perform better in the next epoch (epoch 2). In epoch 2, by implementing the selected policy on the testbed, we collect additional data, which will be added to the dataset, and the same process repeats.

\section{Performance Evaluation}
\label{sec:results}

In this section, we present the results of our O-RAN traffic classifier and PRB optimization framework. We evaluate the classifier in both offline and online settings, demonstrating its deployment as an xApp within the O-RAN architecture. Our evaluation data is generated using the Colosseum testbed, as described in Sec. \ref{sec:dataset}, ensuring a realistic and robust validation of our models.

We organize the performance evaluation into three main subsections. First, we analyze the offline traffic classification accuracy of our proposed models across various temporal input sizes ($T$), highlighting the ability of the classifier to distinguish traffic patterns with high precision under controlled conditions. Next, we assess the online traffic classification accuracy, where the classifier operates as an xApp in a dynamic and resource-constrained environment, validating its real-time performance. Finally, we present results from the PRB optimization framework, demonstrating the impact of reinforcement learning methods on achieving slice-specific performance objectives. This section highlights the interplay between traffic classification and resource allocation, providing a comprehensive view of our system's capabilities.

Through these evaluations, we aim to show that our integrated framework not only achieves high traffic classification accuracy but also effectively allocates PRBs in dynamic 5G environments, meeting diverse SLA requirements across slices. Together, these results validate the feasibility and scalability of our approach for real-world O-RAN deployments.

\subsection{Offline Traffic Classification Evaluation}
\label{sec:results_offline}

\begin{figure}[htb]
    \centering
    \includegraphics[width=.9\linewidth]{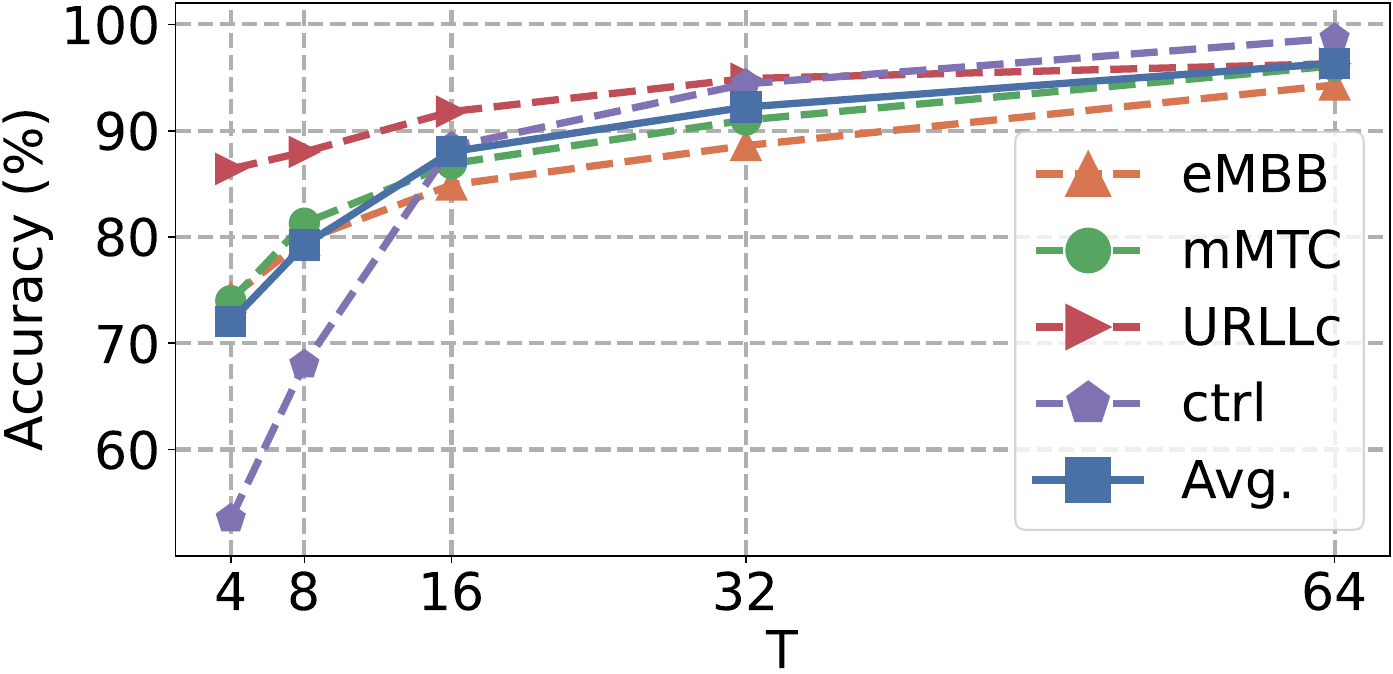}
    \caption{Average offline CNN-based classifier performance for different input slice sizes.}
    \label{fig:CNN_acc}
\end{figure}

In order to test the classifier performance and highlight the challenges of our proposed approach in a realistic O-RAN setting, we train different versions of the model for input slice sizes of $T = \{4, 8, 16, 32, 64\}$, which correspond to classification time granularities of $\{1, 2, 4, 8, 16\}$ seconds, respectively. Fig.~\ref{fig:CNN_acc} shows the CNN-based classification accuracy for different values of $T$, illustrating how longer observation windows improve performance. The classifier achieves its highest accuracy ($\sim 95\%$) when trained with $T=64$, meaning it observes $16$ seconds of sampled KPIs.  

\begin{figure}[htb]
\centering
  \includegraphics[width=\linewidth]{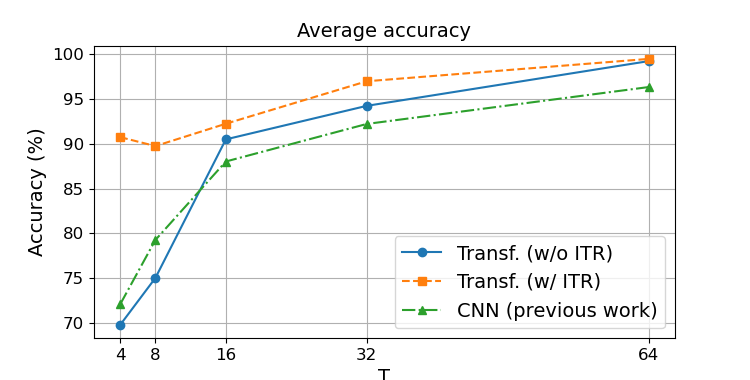}
    \caption{Average offline accuracy of classifiers for different $T = \{ 4, 8, 16, 32, 64\}$. Performance is compared among transformers presented in~\cite{belgiovine2024megatron} with and without the ITR heuristic and CNN model.}
  \label{fig:avg_acc}
\end{figure}

To fairly assess classification accuracy in test settings, we apply the \textit{Idle Traffic Removal (ITR)} heuristic from~\cite{belgiovine2024megatron}. This method filters out idle traffic portions from the incoming KPI stream and adjusts the total number of classification samples, leading to a more accurate performance evaluation.  

% Fig.~\ref{fig:avg_acc} compares the average offline classification accuracy across slices for our CNN and transformer-based models. The ITR heuristic improves accuracy by up to $20\%$ for small window sizes, but has limited impact on the largest window size. This matches intuition. Even for bursty traffic, if you observe long enough, you will see some traffic. On the other hand, if you have a very small window size, you might only observe the periods between a burst. On average, the transformer model outperforms the CNN by approximately $8\%$ in offline evaluations.  

Fig.~\ref{fig:avg_acc} compares the average offline classification accuracy across slices for our CNN and transformer-based models. On average, the transformer model outperforms the CNN by approximately $8\%$ in offline evaluations.  The ITR heuristic improves accuracy by up to $20\%$ for small window sizes but has minimal impact on larger window sizes. This result aligns with intuition: for highly bursty traffic, a longer observation window increases the likelihood of capturing meaningful activity, while very small windows may only capture idle periods between bursts, leading to misclassification. Therefore, removing these short windows composed entirely of \say{silent} portions significantly improves classification accuracy.

\subsection{Online $\TC$ Accuracy Evaluation}
\label{sec:results_online}
We also evaluate the performance of $\TC$ deployed as an xApp in an O-RAN compliant system, shown in Fig.~\ref{fig:system}, to validate the results from the offline trained model. The xApp is deployed inside the near-RT RIC as a Python script, following the implementation details in Sec. \ref{capturing kpis}. The xApp constructs the input to the classifiers by stacking newly received KPIs on the previous $T-1$, and then queries the classifier model in order to provide the classification output. The classifier output identifies which slice the UE should be allocated to at the classification moment and is reported back to the gNB to perform slice assignment.

\begin{figure}[htb]
    \centering
    \includegraphics[width=.9\linewidth]{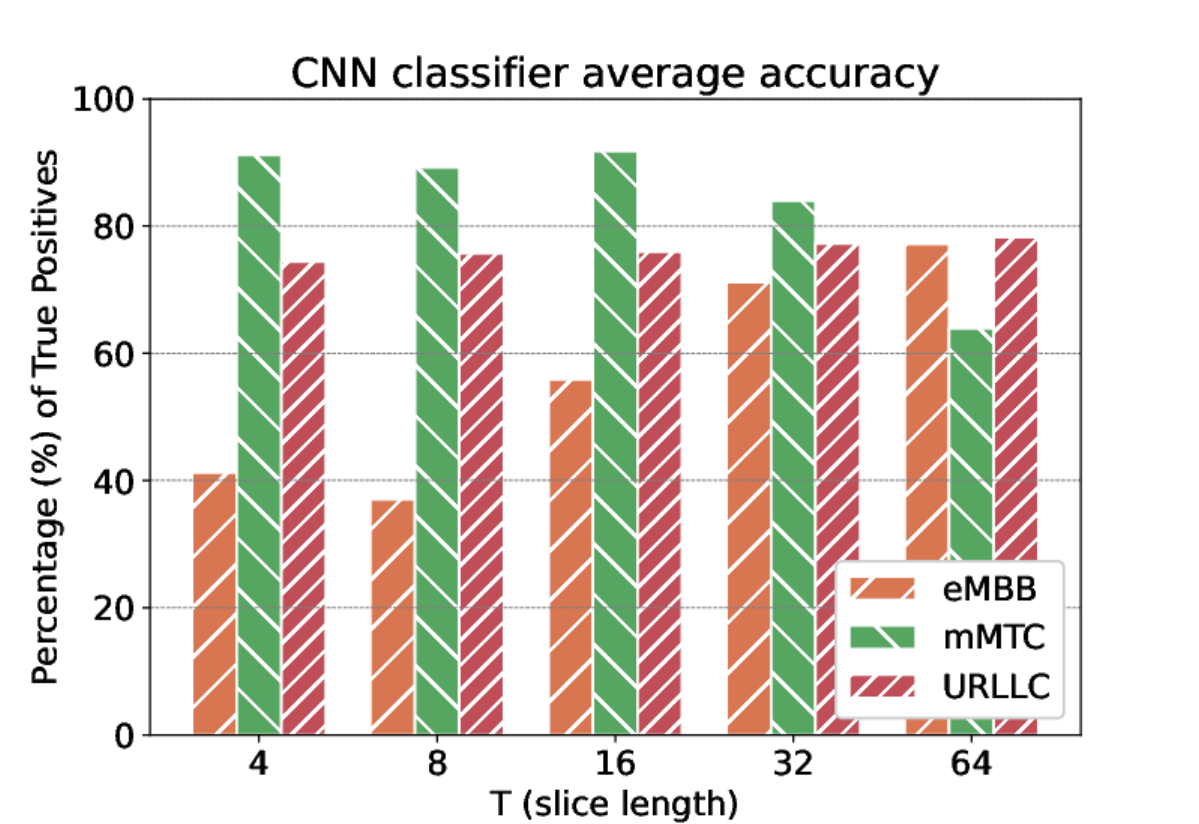}
    \caption{Average accuracy as percentage (\%) of correct classifications for each test in online testing.}
    \label{fig:avg_live_class}
\end{figure}

In order to test the accuracy of our proposed models in unseen conditions, we collect a new set of traffic pattern transmissions for each class (except \textit{ctrl}, which would be naturally included in all the traces) and replay such patterns on Colosseum between the gNB and UE. This generates a new set of KPIs for testing that are different from the KPIs we used for our training. 

We evaluate how the classifier performs on the filtered inputs, again considering all configurations of slice length $T = \{ 4, 8, 16, 32, 64\}$ in Fig. \ref{fig:avg_live_class}. We observe that, while longer context provided by larger values of $T$ generally helps the classification accuracy, there is no one-size-fits-all solution. This highlights the challenges posed by the problem of real time traffic type classification in fully O-RAN compliant systems tackled in this work.

\subsection{RB Optimization}

\begin{figure}[htb]
    \centering
    \includegraphics[width=0.9\linewidth]{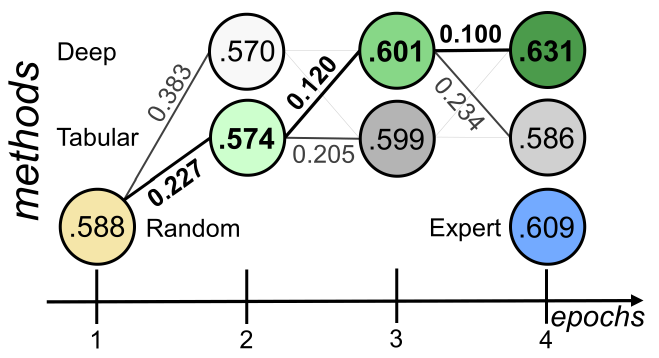}
    \caption{This graph demonstrates how our selection criteria leads to choosing the best policy for each round. The Bellman Error is calculated based on the data in the validation set collected during the current epoch and is displayed as the edge weights. The average performance for the trained model after evaluation is shown in the vertex. After four epochs, the RL policy outperforms the manually configured expert settings.}
    \label{fig:methods_epochs}
\end{figure}

The results of our train-test-improve method with RL algorithms, described in Sec.~\ref{ss:tti} and Sec.~\ref{rl method}, are shown as a graph in Fig.~\ref{fig:methods_epochs}. The process begins with an initial policy that randomly selects an action in epoch 1, represented by the starting node. Each edge weight corresponds to the calculated Bellman Error (BE), as defined in Sec.~\ref{ss:RL_methods}. In a deployed system, we would select the RL policy with the lowest BE for the next round of deployment. To validate this selection process, we deploy and evaluate both RL policies in every new epoch. Each vertex in Fig.~\ref{fig:methods_epochs} represents the average performance of a policy on the common user tuples (listed in Section~\ref{ss:tti}) across all collected data. The performance scores are calculated by averaging the score for each user based on the slice the user belongs to using Equations \ref{eqn:embb}, \ref{eqn:urllc}, and \ref{eqn:mmtc}. For our environment the correct conversion factor is $\kappa = \frac{8}{1\times 10^9}$. We empirically determined the scaling and offsets in Equation~\ref{eqn:embb} based on prior dataset analysis and set them to $\alpha=\frac{1}{3}$, $\beta=\frac{3}{2}$. Finally, we set the maximum RAN latency in seconds as $\gamma = 1$ for Equation~\ref{eqn:urllc}.  

Our results show that the BE metric reliably predicts which policy will perform best in the next epoch, provided that a sufficient amount of data has been collected for performance evaluation. This distinction is critical. For example, in our initial evaluation of epoch 2, the DeepQ method appeared to outperform the TabularQ method (0.624 vs. 0.583). However, a statistical power analysis revealed that both methods had wide confidence intervals (CIs), indicating high variability in their results. When the CI width exceeded 10\% of the mean performance—a predefined threshold for statistical reliability—we computed the number of additional trials required to reduce the CI to an acceptable range. This ensured that the average performance metric accurately reflected model efficacy. After collecting additional data, we found that the TabularQ model used in round 2 actually outperformed the DeepQ model, as shown in Fig.~\ref{fig:methods_epochs}. This result highlights the success of using BE as a selection criterion, even in cases where data scarcity initially leads to misleading performance estimates.

Beyond model selection, Fig.~\ref{fig:methods_epochs} demonstrates the effectiveness of our iterative approach in optimizing PRB configurations. With each epoch, the system identifies improved configurations, progressively enhancing overall performance. By epoch 4, the RL-based policy significantly outperforms a manually configured policy created by an expert, underscoring the advantages of continuous learning in complex, dynamic environments. The iterative refinement process ensures that system performance improves as more data is collected and analyzed. Additionally, the BE selection method consistently identifies the best model for each epoch, reinforcing the system's ability to adapt to a wide range of input parameters. This continuous improvement cycle not only enhances resource allocation efficiency across all slices but also demonstrates the scalability and practical viability of our approach for O-RAN resource management.  

\section{Discussion}\label{discussion}

One of the challenges we identified early on is the vast amount of potential applications and use cases for modern UEs. We chose to use specific exemplar types of traffic, detailed in Sec. \ref{sec:real world 5G}, for our initial exploration. While we do not cover all possible use cases in this work, we believe using select cases of real traffic is a better starting point than using statistical based traffic generation. In general, the majority of internet traffic is inherently bursty and unpredictable, as observed in most of our collected traffic. 

While testing our RL method for PRB optimization, we observed a total of 124 unique user combinations. A subset of these combinations is shown in Table~\ref{tab:mean_performance_by_user}. One notable trend is that certain user combinations consistently achieve higher mean performance scores than others, regardless of the PRB allocation policy used. This suggests that some user distributions inherently present greater challenges for resource allocation.

\begin{table}[htb]
\centering
\rowcolors{2}{gray!20}{}
\begin{tabular}{|l;{1pt/1pt}l;{1pt/1pt}l;{1pt/1pt}l|}
\hline
Users & Mean    & CV      & Trials \\
\hline\hline
0, 1, 0          & 0.9972  & 0.0060  & 9    \\
0, 1, 1          & 0.7830  & 0.0534  & 11   \\
\textbf{0, 2, 2} & \textbf{0.7125}  & \textbf{0.1284}  & \textbf{79}  \\
\textbf{0, 1, 2} & \textbf{0.6731}  & \textbf{0.1049}  & \textbf{92}  \\
\textbf{1, 2, 1} & \textbf{0.6583}  & \textbf{0.0975}  & \textbf{92}  \\
1, 3, 2          & 0.6538  & 0.1460  & 91   \\
\textbf{1, 3, 4} & \textbf{0.5892}  & \textbf{0.1556}  & \textbf{81}  \\
\textbf{1, 2, 3} & \textbf{0.5880}  & \textbf{0.1370}  & \textbf{175} \\
2, 3, 3          & 0.5679  & 0.1399  & 82   \\
\textbf{1, 1, 2} & \textbf{0.5626}  & \textbf{0.1048}  & \textbf{111} \\
\textbf{1, 2, 5} & \textbf{0.5508}  & \textbf{0.1267}  & \textbf{77}  \\
\textbf{1, 1, 4} & \textbf{0.5253}  & \textbf{0.1169}  & \textbf{119} \\
\textbf{3, 2, 3} & \textbf{0.5146}  & \textbf{0.1174}  & \textbf{75}  \\
3, 2, 1          & 0.4947  & 0.1941  & 56   \\
2, 3, 2          & 0.4903  & 0.3782  & 42   \\
3, 0, 0          & 0.1835  & 0.7538  & 6    \\ 
\hline
\end{tabular}
    \caption{Mean performance scores for a subset of user combinations across all policies. The Coefficient of Variation (CV) is an indicator of how consistent the scores are for that user combination. The nine common user configurations are in bold.}
    \label{tab:mean_performance_by_user}
\end{table}

To further analyze performance stability, we calculate the Coefficient of Variation (CV) for each user combination. The CV measures relative variability by normalizing the standard deviation against the mean. A lower CV indicates more consistent performance across trials, while a higher CV suggests greater fluctuations. In Table~\ref{tab:mean_performance_by_user}, we observe that some user combinations exhibit significantly higher variability than others, highlighting that even when the number of users per slice is fixed, traffic demands remain highly dynamic. 

For example, the user combination (2, 3, 2) has a CV of 0.378, reflecting substantial performance variation. Narrowing our focus to the Expert configuration, we observed 11 trials and calculated an even higher CV of 0.792, indicating extreme fluctuations in traffic patterns. This variability suggests that a simple user count per slice may be insufficient for optimal PRB allocation. Incorporating finer-grained KPIs, such as per-user traffic demands, could lead to improved performance. We leave this exploration for future work.

We compare the best-performing policy from epoch 4 to the expert-configured policy for a subset of user combinations in Table~\ref{tab:expert_vs_round4}. The table is divided into two sections: the top four rows represent common user configurations that all models observe in every training round, while the bottom four rows represent less common user configurations, which are not observed by all models in every training round. This distinction allows us to evaluate how well the policy generalizes to both frequently and infrequently seen traffic patterns.

Across these user configurations, the epoch 4 policy outperforms the expert configuration in three out of four cases, achieving higher mean performance for most user combinations. More importantly, the epoch 4 policy exhibits significantly lower variability, as indicated by the lower Coefficient of Variation (CV) in three out of four cases. In some cases, the reduction in CV is by an order of magnitude, suggesting that the learned policy is not only achieving higher performance but also doing so more consistently.

\begin{table}[htb]
\centering
\renewcommand{\arraystretch}{1.2} % Adjust row height for better spacing

\begin{tabular}{|l|r;{1pt/1pt}r|r;{1pt/1pt}r|}
\hline
\multirow{2}{*}{\textbf{Users}} & \multicolumn{2}{c|}{\textbf{Expert}} & \multicolumn{2}{c|}{\textbf{Epoch 4}} \\  
\cline{2-5}
 & \multicolumn{1}{c;{1pt/1pt}}{Mean} & \multicolumn{1}{c|}{CV} & \multicolumn{1}{c;{1pt/1pt}}{Mean} & \multicolumn{1}{c|}{CV} \\  
\hline\hline
0, 1, 2  & \textbf{0.6906}  & 0.1217  & 0.6844  & \textbf{0.0142}  \\
0, 2, 2  & 0.7348  & 0.1362  & \textbf{0.7626}  & \textbf{0.0459}  \\
1, 1, 4  & 0.5421  & 0.0871  & \textbf{0.5684}  & \textbf{0.0303}  \\
1, 2, 3  & 0.6153  & 0.0792  & \textbf{0.6466}  & \textbf{0.0353}  \\
\hline
1, 3, 2  & \textbf{0.7072}  & \textbf{0.0292}  & 0.6986  & 0.0548  \\
2, 3, 3  & 0.5861  & 0.0964  & \textbf{0.6157}  & \textbf{0.0575}  \\
3, 2, 1  & 0.5213  & 0.0567  & \textbf{0.5593}  & \textbf{0.0473}  \\
3, 2, 2  & 0.4941  & \textbf{0.0401}  & \textbf{0.5295}  & 0.0478  \\  
\hline
\end{tabular}

\caption{Comparison of mean performance scores and Coefficient of Variation (CV) between the Expert configuration and the Deep RL policy in epoch 4 for select user combinations. A higher mean indicates better performance, while a lower CV indicates more consistent performance.}
\label{tab:expert_vs_round4}
\end{table}

This is a critical finding because it indicates that the train-test-improve approach successfully learns an optimized PRB allocation strategy, even for user configurations that were not consistently observed in every training round. The ability to generalize and adapt to previously unseen traffic distributions suggests that the model is not simply memorizing fixed allocation patterns but is instead capturing underlying resource allocation principles that improve performance across a broad range of scenarios. This result reinforces the effectiveness of reinforcement learning for dynamic PRB optimization in O-RAN systems, demonstrating that iterative training yields more robust and adaptive policies compared to static, expert-designed configurations.

% For the train-test-improve pipeline, we adopted a policy selection method aimed at maximizing performance in the next round. However, it is also important to consider the exploration aspect of a policy, as exploration can enhance future learning by incorporating new data. The challenge lies in quantifying a policy's exploration capability and balancing performance with exploration in each round.

In this paper, we adopt a policy selection method in the train-test-improve pipeline that prioritizes maximizing performance in the next round. However, it is also important to consider the exploration aspect of a policy, as exploration can enhance future learning by incorporating new data. The challenge lies in quantifying a policy's exploration capability and balancing immediate performance with long-term learning. While our framework supports using multiple policies and selection methods, we leave the specific implementation of alternative selection strategies for future work.

\section{Conclusion}
\label{sec:challenges}

To meet the complex demands of 5G and beyond cellular networks, we introduce \textbf{TRACTOR+} (\textbf{Tr}affic \textbf{A}nalysis and \textbf{C}lassification \textbf{T}ool for \textbf{O}pen \textbf{R}AN \textbf{+}PRB optimization), showcasing the viability of automated traffic classification and per slice PRB allocation optimization. We offer public access to our tool-set, along with O-RAN compliant KPIs generated from a real-user 5G traffic dataset. Initial ML models validate TRACTOR's capacity for automated user traffic classification, achieving accuracy of up to $99\%$ offline and $>92\%$ online. Additionally, we extend TRACTOR’s capabilities with reinforcement learning-based PRB allocation, demonstrating that the framework can dynamically optimize resource distribution across network slices to improve performance. These contributions highlight TRACTOR+’s flexibility as a research platform for advancing ML-driven network optimization.

Future research in this space will need to address several key challenges. There are additional opportunities to optimize PRB allocation, particularly in adapting to real-time traffic fluctuations and balancing resource efficiency with performance guarantees. More sophisticated ML techniques could improve slice management by predicting traffic patterns in advance, allowing networks to proactively allocate resources rather than reacting to congestion as it occurs. The timescale of KPI reporting also presents a trade-off between responsiveness and computational overhead, requiring careful consideration of how frequently network conditions should be assessed. Additionally, as O-RAN systems evolve, further exploration is needed to understand the impact of multi-user interactions, competing traffic demands, and the scalability of ML-driven policies in distributed large-scale deployments.

This work underscores the potential for future networks to excel across multiple performance metrics while maintaining adaptability, efficiency, and scalability. By enabling real-world experimentation with ML-driven traffic classification and dynamic resource allocation, TRACTOR+ provides a powerful and flexible framework for ongoing research in O-RAN systems.

\section*{Acknowledgment}
This article is based upon work partially supported by U.S.\ National Science Foundation under grants ECCS-2228974, CNS-2120447, CNS-2112471, and CNS-1925601.

\footnotesize  % for natbib
\bibliographystyle{IEEEtran}
\bibliography{IEEEabrv,references}

\begin{IEEEbiography}[{\includegraphics[width=1in,height=1.25in,clip,keepaspectratio]{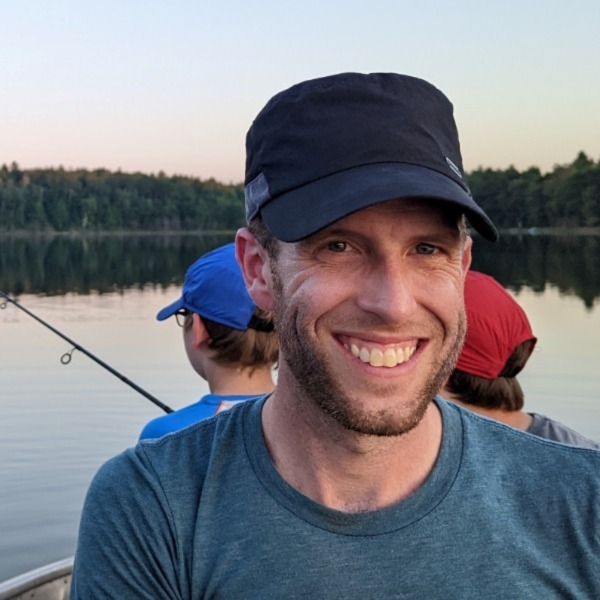}}]%
{Joshua Groen} received the BSE degree in electrical
engineering from Arizona State University, in 2007,
and the MS degree in electrical engineering from the
University of Wisconsin, in 2017. He is currently
working toward the PhD degree with Northeastern
University. His research interests include wireless communications, security, and machine learning. Previously he worked with the US Army
Regional Cyber Center – Korea as the senior network
engineer. 
\end{IEEEbiography}

\vspace{-0.5cm}
\begin{IEEEbiography}[{\includegraphics[width=1in,height=1.25in,clip,keepaspectratio]{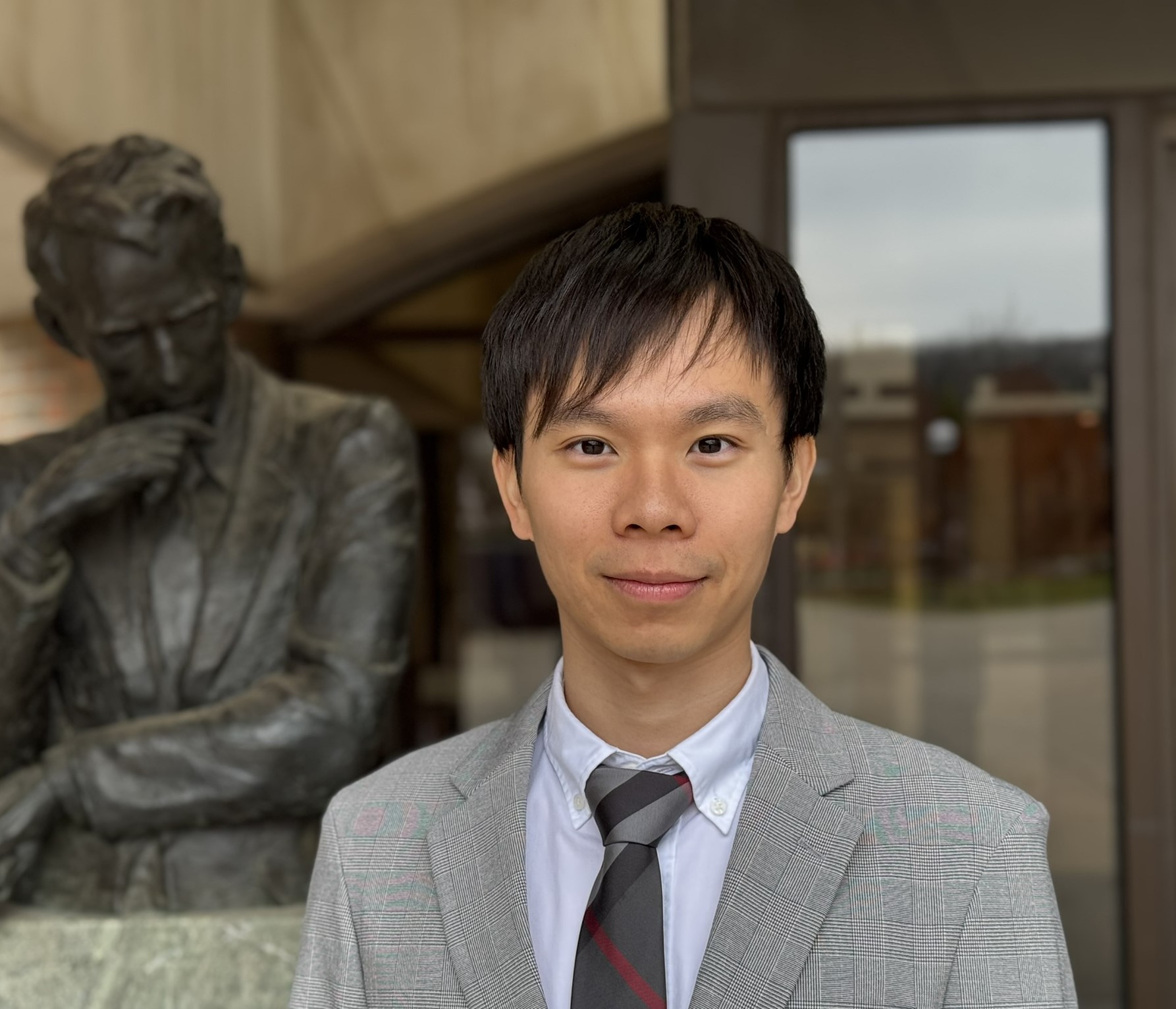}}]%
{Zixian Yang} is a postdoctoral research fellow in the Electrical Engineering and Computer Science Department at the University of Michigan, Ann Arbor, advised by Prof. Lei Ying. Previously, Zixian completed his Ph.D. in the same department in 2025. His research interests lie in joint online learning and decision making problems, which include communications, recommendation, queueing, scheduling, matching, and pricing in unknown environments.
\end{IEEEbiography}

\vspace{-0.5cm}
\begin{IEEEbiography}[{\includegraphics[width=1in,height=1.25in,clip,keepaspectratio]{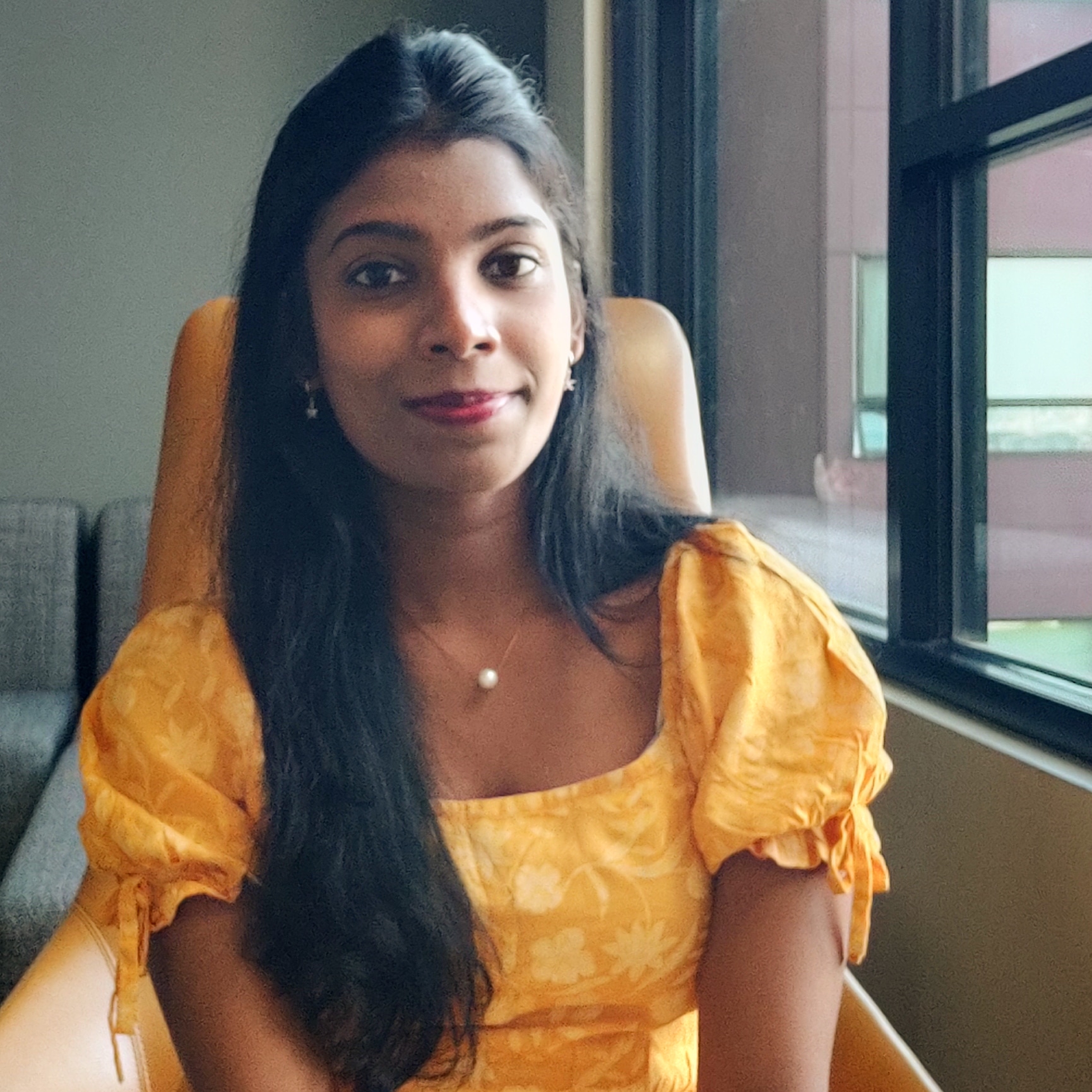}}]%
{Divyadharshini Muruganandham} is currently pursuing a Ph.D. degree in computer engineering at University of Texas at Austin under the supervision of Prof. K. Chowdhury. Her current
research focuses on applied AI/ML in wireless communication, mmWave beamforming, network optimization,
and actively involved in experimental deployments of emerging wireless technologies, aimed at shaping the future of next-generation communication systems.
\end{IEEEbiography}

\vspace{-0.5cm}
\begin{IEEEbiography}[{\includegraphics[width=1in,height=1.25in,clip,keepaspectratio]{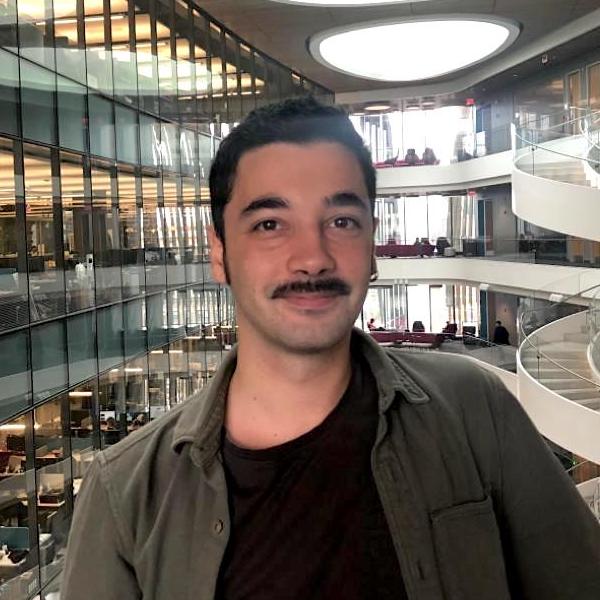}}]%
{Mauro Belgiovine} is pursuing his Ph.D. at the Electrical and Computer Engineering Department at Northeastern University, Boston, Massachusetts, under the guidance of Prof. Kaushik Chowdhury. His current research interests involve deep learning, wireless communication, heterogeneous computing and digital twins.
\end{IEEEbiography}

\vspace{-0.5cm}
\begin{IEEEbiography}[{\includegraphics[width=1in,height=1.25in,clip,keepaspectratio]{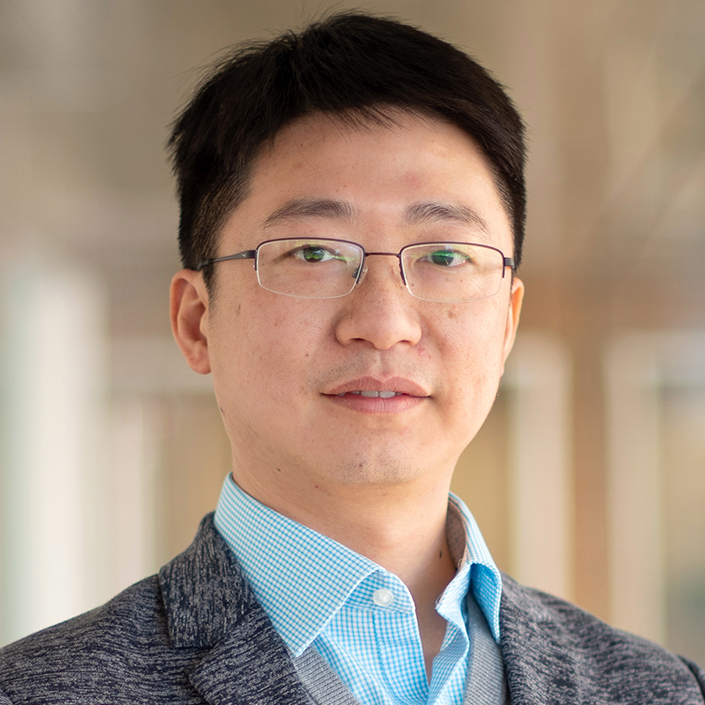}}]%
{Lei Ying} received his B.E. degree from Tsinghua University, Beijing, China, and his M.S. and Ph.D. in Electrical and Computer Engineering from the University of Illinois at Urbana-Champaign. He currently is a Professor at the Electrical Engineering and Computer Science Department of the University of Michigan, Ann Arbor, and an IEEE Fellow. His research is broadly in the interplay of complex stochastic systems and big data, including reinforcement learning, large-scale communication/computing systems for big-data processing, private data marketplaces, and large-scale graph mining. 
\end{IEEEbiography}

\vspace{-0.5cm}
\begin{IEEEbiography}[{\includegraphics[width=1in,height=1.25in,clip,keepaspectratio]{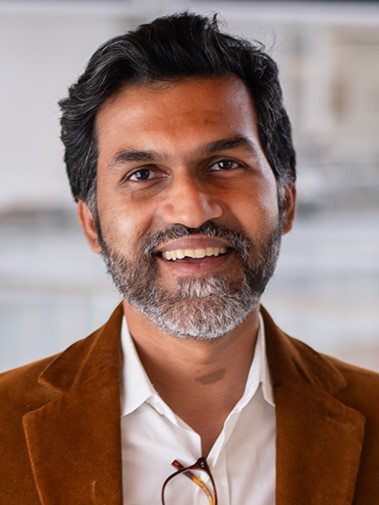}}]%
{Kaushik Chowdhury} received the PhD degree from
the Georgia Institute of Technology, in 2009. He is a Chandra Family Endowed Distinguished Professor in Electrical and Computer Engineering at The University of Texas at Austin. His current research interests involve systems aspects of machine learning for agile spectrum
sensing/access, unmanned autonomous systems, programmable and open cellular networks, and largescale experimental deployment of emerging wireless
technologies.
\end{IEEEbiography}
\end{document}